\documentclass[twocolumn,secnumarabic,amssymb, nobibnotes, aps, prb, superscriptaddress]{revtex4-2}
\usepackage{color, soul}
\usepackage{graphicx}
\usepackage{float}
\usepackage{amsmath}
\RequirePackage[hyperfigures=true, hyperfootnotes=true, hyperindex=true,breaklinks=true]{hyperref}
\definecolor{link_green}{rgb}{0.0,0.7,0.0}
\definecolor{link_blue_dark}{rgb}{0.0,0.0,0.7}
\definecolor{link_blue}{rgb}{0.0,0.0,1}
\definecolor{link_red}{rgb}{0.7,0.0,0.0}
\definecolor{link_red_dark}{rgb}{0.6,0.0,0.0}
\definecolor{link_red_verydark}{rgb}{0.3,0.0,0.0}
\hypersetup{hyperfigures=true,hyperfootnotes=true,hyperindex=true,pagebackref,breaklinks=true,citecolor=link_green,filecolor=link_blue,linkcolor=link_blue_dark, urlcolor=link_blue,colorlinks,bookmarks,bookmarksopen,bookmarksnumbered, pdfauthor="Olivier Fruchart",pdfstartview=FitH} \RequirePackage[all]{hypcap}

\usepackage{siunitx}%
\RequirePackage{xspace}
\RequirePackage{upgreek}%
\RequirePackage{journames}

  \RequirePackage[normalem]{ulem} 
  \RequirePackage{soul}
  \definecolor{olivier_comment}{rgb}{0.6,0.6,0.9}
  \definecolor{olivier_add}{rgb}{0.8,0.2,0.2}
  \definecolor{olivier_remove}{rgb}{0.5,0.5,0.5}

\RequirePackage[mathcal]{euscript}

\def\eg{{\it e.g.}\/\xspace}%
\def\ie{{\it i.e.}\/\xspace}%
\def\etal{{\it et~al.}\/\xspace}%
\newcommand{\vect}[1]{\mathbf{#1}}
\newcommand{\vectNabla}{\boldsymbol{\nabla}}
\def\muZero{\mu_{\mathrm 0}}%
\def\Ms{M_{\mathrm s}}
\def\vectHd{\vect{H}_{\mathrm d}}
\def\vectHeff{\vect{H}_\mathrm{eff}}
\def\vectM{\vect M}
\def\vecthd{\vect{h}_{\mathrm d}}
\def\vectm{\vect m}
\def\vectj{\vect j}
\def\vectHOE{\vect{H}_\mathrm{\OE}}
\newcommand{\vecthat}[1]{\boldsymbol{\hat{\vect{#1}}}}
\newcommand{\unitvect}[1]{\vecthat{#1}}
\newcommand{\unitvecttheta}{\hat{\boldsymbol{\uptheta}}}
\newcommand{\unitvectvarphi}{\hat{\boldsymbol{\upvarphi}}}
\def\Kd{K_{\mathrm d}}
\newcommand{\OErsted}{\OE rsted\/\xspace}%
\newcommand{\lex}{\ell_\mathrm{ex}}
\newcommand{\muB}{\mu_\mathrm{B}}
\newcommand{\DeltaT}{\Delta_\mathrm{T}}
\newcommand{\jc}{j_\mathrm{c}}
\newcommand{\tauc}{\tau_\mathrm{c}}
\newcommand{\secref}[1]{\hbox{sec.\ref{#1}}}
\newcommand{\bracketsecref}[1]{~(\secref{#1})}%

\definecolor{arnaud_comment}{rgb}{0.6,0.6,0.9}
\definecolor{arnaud_add}{rgb}{0.0,0.5,1}
\definecolor{arnaud_remove}{rgb}{0.5,0,0}

\newcommand{\ve}[1]{\mathbf{#1}}

\begin{document}

\title{Mechanism of current-assisted Bloch-point wall stabilization for ultra fast dynamics}

\author{A. De Riz}
\email{arnaud.deriz@cea.fr}
\affiliation{Univ. Grenoble Alpes, CNRS, CEA, Spintec, Grenoble, France}
\author{J.~Hurst}
\affiliation{Univ. Grenoble Alpes, CNRS, CEA, Spintec, Grenoble, France}
\author{M.~Schöbitz}
\affiliation{Univ. Grenoble Alpes, CNRS, CEA, Spintec, Grenoble, France}
\affiliation{Friedrich-Alexander Univ. Erlangen-Nürnberg, Inorganic Chemistry, Erlangen, Germany}
\author{C.~Thirion}
\affiliation{Univ. Grenoble Alpes, CNRS, Institut Néel, Grenoble, France}
\author{J.~Bachmann}
\affiliation{Friedrich-Alexander Univ. Erlangen-Nürnberg, Inorganic Chemistry, Erlangen, Germany}
\affiliation{Saint-Petersburg State University, Institute of Chemistry, Universitetskii pr.~26, 198504, St.~Petersburg, Russia}
\author{J.C.~Toussaint}
\affiliation{Univ. Grenoble Alpes, CNRS, Institut Néel, Grenoble, France}
\author{O.~Fruchart}
\affiliation{Univ. Grenoble Alpes, CNRS, CEA, Spintec, Grenoble, France}
\author{D.~Gusakova}%
\email{daria.gusakova@cea.fr}
\affiliation{Univ. Grenoble Alpes, CNRS, CEA, Spintec, Grenoble, France}

\date{\today}%

\begin{abstract}
Two types of domain walls exist in magnetically soft cylindrical nanowires: the transverse-vortex wall~(TVW) and the Bloch-point wall~(BPW). The latter is expected to prevent the usual Walker breakdown, and thus enable high domain wall speed. We showed recently [M.~Schöbitz \etal, Phys. Rev. Lett. 123, 217201 (2019)] that the previously overlooked \OErsted field associated with an electric current is a key in experiments to stabilize the BPW and reach speed above \SI{600}{\meter\per\second} with spin-transfer. Here, we investigate in detail this situation with micromagnetic simulations and modeling. The switching of the azimuthal circulation of the BPW to match that of the \OErsted field occurs above a threshold current scaling with~$1/R^3$~($R$ is the wire radius), through mechanisms that may involve the nucleation and/or annihilation of Bloch points. The domain wall dynamics then remains of a below-Walker type, with speed largely determined by spin-transfer torque alone.

\end{abstract}

\maketitle

\section{Introduction}

Domain wall motion has been a crucial aspect in the study of the magnetization dynamics since the early days of the understanding of coercivity\cite{bib-KON1937}. One initially considered motion under an applied magnetic field\cite{bib-THI1973,bib-ONO1999}, extended since the early 2000's to current-driven cases, based on spin-transfer effects\cite{bib-GRO2003b} and more recently spin-orbit torques\cite{bib-MIR2010} and other effects: heat gradients\cite{bib-CHA2010}, strain\cite{bib-DEA2015}, spin waves\cite{bib-HIN2011} etc. As such, the analysis of domain wall motion is a powerful probe of condensed matter magnetism phenomena, \eg, allowing one to evaluate adiabatic versus non-adiabatic spin-transfer torques\cite{bib-MIR2009b}, the strength of a Dzyaloshinskii-Moriya interaction\cite{bib-THI2012} etc.

The consideration of one-dimensional conduits such as nanostrips or cylindrical nanowires for domain wall motion, provides a situation with low complexity, suitable for a reliable analysis, and thereby represents a textbook case. Such conduits also offer the prospect for memory and logic devices\cite{bib-ALL2005}, and more recently for neuromorphic computing\cite{bib-LEQ2016}. Domain wall motion is intrinsically related to the precessional dynamics of magnetization, described by the Landau-Lifshitz-Gilbert(LLG)-Slonczewski equation\cite{bib-THI2005}, see Eq.\ref{eq:LLG} later on. A common feature of motion under both field and current is the steady-state propagation under low stimulus, and a precessional regime above a threshold, with a cross-over process called the Walker breakdown. In the former regime, an internal restoring force gives rise to torques that balance the ones responsible for azimuthal precessional, while also possibly contributing to motion. Common sources of restoring forces include dipolar energy (\eg, the shape anisotropy of a thin film with in-plane magnetization) and the Dzyaloshinskii-Moriya interaction. It is the finite and sometimes moderate magnitude of these restoring forces, that cannot balance large driving stimulus, which leads to the Walker breakdown and oscillations or chaotic changes of internal degrees of freedom in domain walls.

Contrary to most cases of thin films, the absence of Walker breakdown was predicted for head-to-head domain walls in magnetically soft cylindrical nanowires, made possible by the existence of a specific domain wall called the Bloch-point Wall~(BPW)\cite{bib-FOR2002b,bib-HER2002a,bib-THI2006}. In the BPW magnetization is curling along the perimeter of the wire to best close magnetic flux~[Fig.\ref{fig:schematics}], creating boundary conditions that impose the existence of a micromagnetic singularity on the wire   axis at the core of the wall, called the Bloch point\cite{bib-FEL1965,bib-DOE1968}. The BPW is of lowest energy for wire diameter above circa seven times the dipolar exchange length $\lex = \sqrt{2A/(\muZero\Ms^2)}$\cite{bib-FOR2002b,bib-THI2006}, with $A$ the exchange stiffness, $\muZero$ the magnetic permeability, and $M_\mathrm{s}$ the spontaneous magnetization. Smaller wire diameters favor the thin-film like transverse-vortex wall, TVW\cite{bib-FRU2015b}. Important for the present work is to mention that two cases of BPW exist, with opposite signs of curling~(also called circulation), which are degenerate at rest. Upon motion the one positive with the direction of motion tends to be favored. The Walker breakdown does not occur, as it would require a too large energy of the dipolar origin with a head-on magnetization configuration along all three directions.

The existence of the BPW was confirmed experimentally at rest in 2014\cite{bib-FRU2014}. However, the first report of its motion under magnetic field was disappointing\cite{bib-FRU2019}: unexpectedly, a change of topology occurs between the BPW and the TVW, liable for instabilities and low speed. More recently, however, we showed that the situation is drastically different for motion driven with a spin-polarized current\cite{bib-FRU2019b}. BPWs remain stable and with speed exceeding $\SI{600}{\meter\per\second}$, setting an experimental record for a purely spin-transfer-driven case. In that work, we showed by simulation that the reason for the robustness of the BPW is its stabilization by the azimuthal \OErsted field, an ingredient disregarded in previous simulations. The resulting circulation is left-handed with respect to the direction of motion, e.g. is opposite to that previously expected from the motion and the chirality of the LLG equation\cite{bib-THI2006}, and is consistent with our experiments.

Thus, the \OErsted field seems crucial for understanding the unusually-high speed of BPWs experimentally, and as such, deserves a thorough investigation. The purpose of the present manuscript is to provide a comprehensive picture of the effect of the \OErsted field on domain walls in cylindrical nanowires, to set the ground for future experimental investigations, and for example guide the search for the magnonic regime occurring around $\SI{1}{\kilo\meter\per\second}$, also called the spin-Cherenkov effect\cite{bib-YAN2011b}. We first present the numerical methods used for the work\bracketsecref{sec-micromagneticMethods}, then examine the processes involved in the stabilization and selection of a specific circulation of the BPW\bracketsecref{sec-selectionOfCirculation}, and finally revisit the expected speed of BPWs under both spin-transfer and \OErsted field. In the manuscript we present results with units of length and current normalized with micromagnetic quantities, making the present work scalable to any magnetically soft material.

\section{Micromagnetic methods}
\label{sec-micromagneticMethods}

\subsection{Micromagnetic equations}

\begin{figure}[t]
   \includegraphics{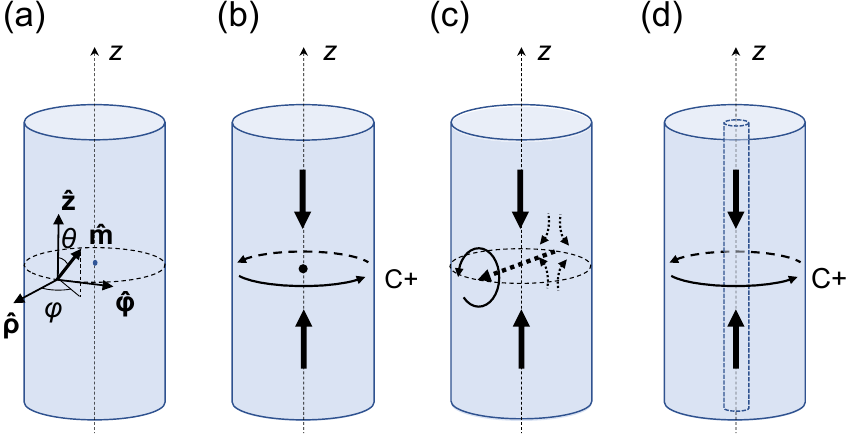}
   \caption{\label{fig:schematics} Schematic of (a) a nanowire, (b) a head-to-head Bloch-point Wall with a positive circulation versus $\unitvect{z}$, (c) a head-to-head transverse-vortex wall, (d) a thick-walled tube with a pseudo-Bloch-point wall.}
\end{figure}

We performed micromagnetic simulations using our finite-element-based software FeeLLGood \cite{bib-ALO2006,bib-ALO2012,bib-KRI2014,bib-FEE}, which solves the Landau-Lifshitz-Gilbert~(LLG)-Slonczewski equation, taking into account the effect of spin transfer\cite{bib-THI2005}:
%
\begin{equation}
\begin{split}\label{eq:LLG}
\partial_t \textbf{m} = &-\gamma_0(\ve{m}\times \vectHeff)+\alpha(\ve{m}\times\partial_t\ve{m})\\
& -(\ve{u}\cdot\nabla)\ve{m}+\beta\ve{m}\times(\ve{u}\cdot\nabla)\ve{m},
\end{split}
\end{equation}
where $\textbf{m}=\textbf{M}/\Ms$ is the magnetization unit vector, $\gamma_0=\muZero|\gamma|$ with $\gamma$ the gyromagnetic ratio, $\vectHeff$ the effective field, $\alpha$ the Gilbert damping and $\beta$ the non-adiabatic coefficient. $\textbf{u} = -g\muB P\vect{j}/(2e\Ms)$ is the velocity field, with $g$ the Landé factor, $\muB$ the Bohr magneton, $P$ the polarization ratio of the spins of flowing conduction electrons, $e$ the elementary charge and $\vectj$ the electric current density. As usual, the positive current direction opposes the electron flow. The effective field is derived from the energy of the system. In this work we consider only the exchange energy, the magnetostatic energy and the Zeeman energy due to the \OErsted field generated by the applied current.

\subsection{Material and computation parameters}
The system of interest is a straight cylindrical nanowire with radius~$R$ made of a magnetically soft material: permalloy  $\mathrm{Fe}_{20}\mathrm{Ni}_{80}$  (exchange stiffness $A = \SI{1e-11}{\joule\per\meter}$,  $\Ms= \SI{8e5}{\ampere\per\meter}$) or $\mathrm{Co}_{20}\mathrm{Ni}_{80}$ ($A =\SI{1.1e-11}{\joule\per\meter}$, $\Ms= \SI{6.7e5}{\ampere\per\meter}$). We consider $P = 0.7$ for both materials. Thanks to the usual micromagnetic normalization of fields to $\Ms$ and lengths to the dipolar exchange length $\lex$, the present results may be scaled to any magnetically soft material~(see Appendix~\ref{sec-appendixDimensionless}). $\lex \simeq\SI{5}{\nano\meter}$ for permalloy, and $\SI{6.25}{\nano\meter}$ for $\mathrm{Co}_{20}\mathrm{Ni}_{80}$.

The mesh is composed of tetrahedrons with characteristic size $\SI{4}{\nano\meter}$, chosen to be slightly smaller than~$\lex$. The surface magnetic charges at wire ends are disconsidered, in order to mimic an infinite wire. We consider an instantaneously applied uniform and steady spin-polarized charge current flowing along the wire axis~$z$.

\subsection{The thick-walled tube ansatz}
\label{sec:thick-walled}

\begin{figure}[t]
\includegraphics[width=8cm]{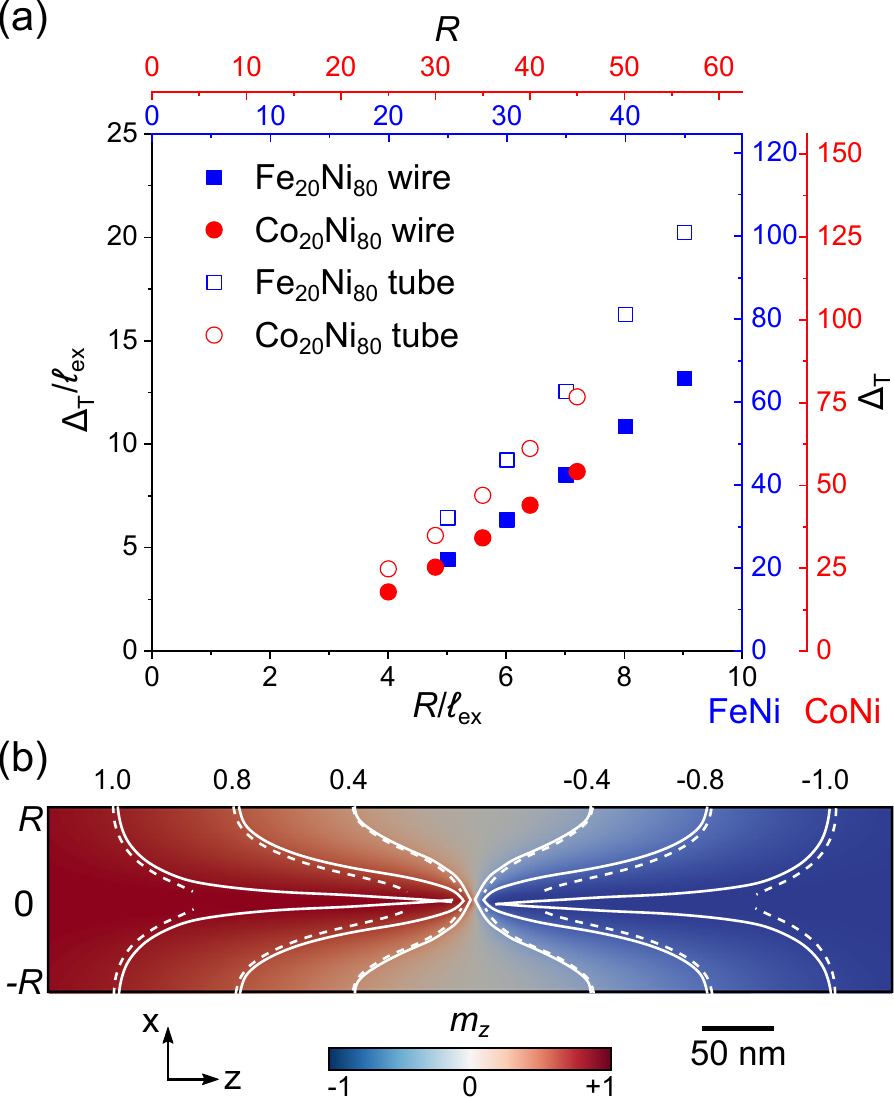}
\caption{\label{fig:static_slice_thielevsR} (a) Domain wall width (Thiele definition) versus the external radius, expressed in real length~(top and right axis) and normalized with the dipolar exchange length $\lex$~(bottom and left axis) (b) Section in the $xz$ plane of a wire. The color map represents the $z$ component of magnetization. The contour lines represent isovalues of $m_z$. The solid lines correspond to the wire, the dotted lines to the tube.}
\end{figure}

A key hypothesis of the micromagnetic theory\cite{bib-BRO1963b} is the description of magnetization with a continuous vector field of uniform and constant modulus. It is therefore not suitable mathematically to describe a Bloch point, involving a singularity in the vector field. In numerical micromagnetism involving Bloch points, this mismatch induces artefacts like the pinning on the discrete numerical lattice during magnetization dynamics, or the logarithmic convergence of magnetization processes such as nucleation\cite{bib-THI2003}. For example, in the finite element approach, the constraint on the magnetization norm is imposed at the mesh nodes. Within every volume element magnetization is interpolated linearly, with its norm possibly greatly reduced, allowing a magnetic object resembling a Bloch point to be centered inside these. This magnetic object may move from one volume element to a neighboring one, however over an energy barrier, inducing a numerical frictional force that depends on the mesh size\cite{bib-HER2004}.

An atomistic model obviously provides an improvement, the mesh being scaled down to the ultimate size of atoms. However, due to the logarithmic convergence mentioned above, an intrinsic pinning remains on the lattice, of the order of a few~mT\cite{bib-AND2014b}. This effect may be responsible for the excitation of helical instabilities sometimes evidenced during motion of the BPW under a large driving force\cite{bib-HER2015}. The question may arise, to which extent this reflects experimental physics. Indeed, in the atomistic models implemented so far, the magnetic moments maintain a fixed magnitude on every lattice site. This is not realistic for band magnetism such as for Fe, Co, Ni and their alloys, for which one expects a local reduction of band splitting and thus atomic moment, allowing to reduce the total energy of the system\cite{bib-BLUe2007}.

The Landau-Lifshitz-Bloch~(LLB) formalism aims at describing such situations, allowing for a spatial variation of the magnetization modulus by introducing a longitudinal susceptibility\cite{bib-GAR1997b}. The Bloch point has been described by a LLB model, down to a cell size of $\SI{0.5}{\nano\meter}$\cite{bib-LEB2012}, however the impact on pinning has not been evaluated. Also, from a fundamental point of view, it is not clear to which extent the fitting of LLB parameters to macroscopic quantities such as the Curie temperature, adequately reflects sub-nm physics with strong gradients of magnetization in the case of band magnetism.

Thus, at this stage we consider that it remains an open question, to which extent Bloch points may be described suitably by simulation, especially regarding their motion. So, in the course of the present report we sometimes consider and compare two situations: that of a wire, and that of a thick-walled tube (\ie, a wire with an empty core of very small radius, \SI{5}{\nano\meter}). Strictly speaking there is no more Bloch point in a thick-walled tube for a BPW at rest, so that in the manuscript we refer to the wall as pseudo-Bloch-point wall~(PBPW). This type of wall in a nanotube is often called a vortex wall, in the literature.

Qualitatively, the physics of domain walls in wires and tubes indeed display many similarities, such as the possible absence of Walker breakdown and the magnonic regime\cite{bib-THI2006,bib-YAN2011b}. Quantitatively, features of tubes tend to converge to wires when the thickness of the tubes is increased\cite{bib-LAN2010}. Below we compare the two situations at equilibrium, to provide a basis for our ansatz.

We characterized both walls through their width, following the Thiele definition:
\begin{equation}\label{eq:thiele}
\DeltaT = \frac{2S}{\int (\partial_z \textbf{m})^2\mathrm{d} V}\;,
\end{equation}
where $S$ is the section of the nanowire or nanotube. Fig.\ref{fig:static_slice_thielevsR}(a) shows that $\DeltaT$ increases with radius $R$ for both situations. This graph is plotted with lengths scaled to $\lex$, to provide a material-independent curve. First, note that while different materials fall on the very same curve for wires, a slight shift exists for tubes. This arises as through normalizing with $\lex$, the inner radius $\SI{5}{\nano\meter}$ converts in a slightly different geometry upon normalization. However, the main point in this plot is the sizable difference of width between a BPW and a PBPW, although the difference in section $S$ is only a few percent. To understand this, we examine the micromagnetic distribution of both walls in the $xz$ plane~[Fig.\ref{fig:static_slice_thielevsR}(b)]. While the two walls share a similar configuration near the outer surface, they differ significantly close to the axis. The absence of Bloch point in the tube removes the need for the pinching of magnetization, explaining a larger width there. It is because the Thiele definition puts a larger weight on locations with a large magnetization gradient, that the resulting width is significantly different although the volume with significant differences is rather small. The similarity of the two maps of magnetization on the outer part of the structure, where the \OErsted field driving the dynamics is largest, makes us confident that a thick-walled tube is a reasonable ansatz for a wire. Note that the \OErsted field used in the following is the one consistent with a tube:
\begin{equation}
\vectHOE(\rho) = \frac{\ve{j}\rho}{2}\left(1-\frac{R_i^2}{\rho^2}\right)\;,
\end{equation}
with $\rho$ is the radial coordinate, and $R_i$ is the inner radius.


\section{Stabilization of the Bloch-point wall and selection of its circulation by the \OErsted field}
\label{sec-selectionOfCirculation}

In this section we examine in detail the role of the \OErsted field on selecting the type of domain wall in nanowires, which we reported only briefly upon its discovery\cite{bib-FRU2019b}. We first show how TVWs are converted in BPWs, and then, how the circulation of a BPW may switch under an \OErsted field of opposite circulation: its phenomenological description, its microscopic and topological understanding, and the threshold of current required for the selection.

\subsection{Transformation of the TVW into a BPW}
\label{sec-selectionOfCirculationTW}

Experimentally, magnetically soft nanowires exhibit domain walls of both TVW and BPW types, in the as-grown state as well as following ac demagnetization along a direction transverse to their axis\cite{bib-BIZ2013,bib-FRU2014,bib-IVA2016b}, or motion of walls under an axial magnetic field\cite{bib-FRU2019}. On the contrary, only BPWs are observed following the application of pulses of current\cite{bib-FRU2019b}, which is not explained by considering the effect of spin transfer alone\cite{bib-WIE2010}.

To understand this, we simulated the response of TVWs in $\mathrm{Co}_{20}\mathrm{Ni}_{80}$ nanowires subject to current pulses, taking into account the effect of the resulting \OErsted field only. We evidence the existence of a threshold current, below which the structure of the TVW is only deformed, while above it the TVW is converted to a BPW. The latter occurs through the peripheral motion of the surface vortex and antivortex~[Fig.\ref{fig:schematics}(c)] towards each other, until they merge, nucleating a Bloch point that then moves radially towards the axis, ending in a BPW. The BPW then reaches a steady configuration under the current pulse, and remains after removal of the \OErsted field, with a circulation positive with respect to the direction of applied current~$\vect j$. This process is similar in topology with the dynamical transformation of a TVW subject to a longitudinal magnetic field\cite{bib-FRU2019}. Qualitatively, in the present case the transformation can be understood as the BPW  and the \OErsted field share the same azimuthal symmetry, thus lowering the energy of the system against a TVW.

For a diameter of $\SI{90}{\nano\meter}$, our simulations point at a threshold current for the TVW-BPW transformation of $\SI{2.8e11}{\ampere\per\meter\squared}$. This value occurs not to be very dependent on the wire diameter, at least for diameters in the range $\SI{75}{\nano\meter}$-$\SI{95}{\nano\meter}$. This low threshold explains why in Ref.\cite{bib-FRU2019b} one observes only BPWs after current pulses, whose magnitude was around $\SI{e12}{\ampere\per\meter\squared}$, suitable for the spin-transfer-torque-driven motion of domain walls. 


\subsection{Switching of circulation of the BPW: phenomenology}

\begin{figure}[t]
\centering\includegraphics{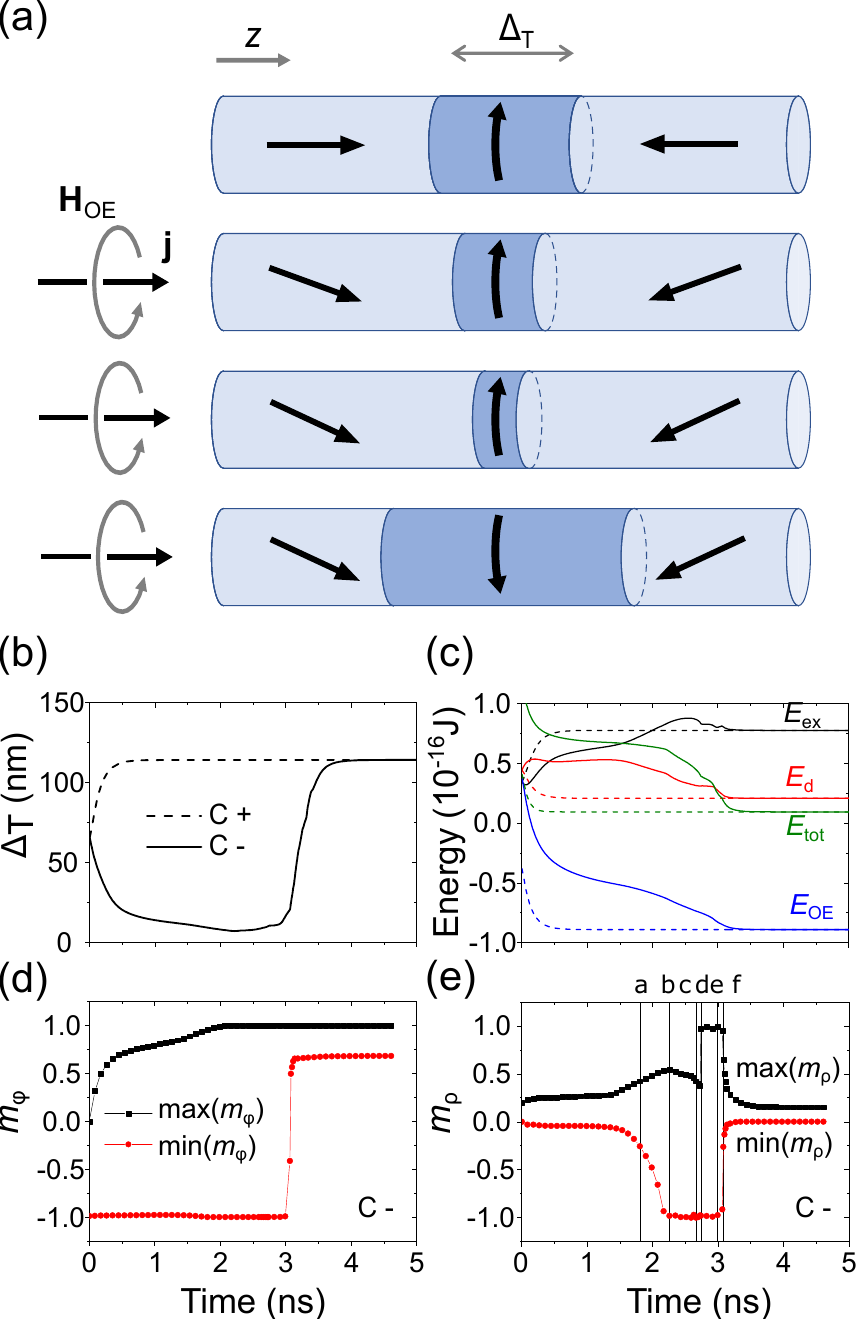}
\caption{\label{fig:sw_thiele_energies_vs_time}(a) Schematic of the circulation switching for a BPW initially $C-$. (b)-(e) Time evolution of four quantities illustrating the response of BPWs under the \OErsted field in a permalloy wire with diameter $\SI{90}{\nano\meter}$ and the applied current density $\SI{1.2e12}{\ampere\per\meter\squared}$, corresponding to an \OErsted field of $\SI{34}{\milli\tesla}$ at the external surface: (b)~the Thiele wall parameter, (c)~micromagnetic energies, and the maxima and minima of the (d)~azimuthal and (e)~radial components of magnetization at the surface of the wire. The a, b, c, d, e, f labels correspond to timestamps of Fig.\ref{fig:sw_maps_mecanism}. The solid line stands for initially negative circulation $C-$ and dashed line for initially positive positive circulation $C+$.}
\end{figure}

\begin{figure*}[t]
\centering\includegraphics{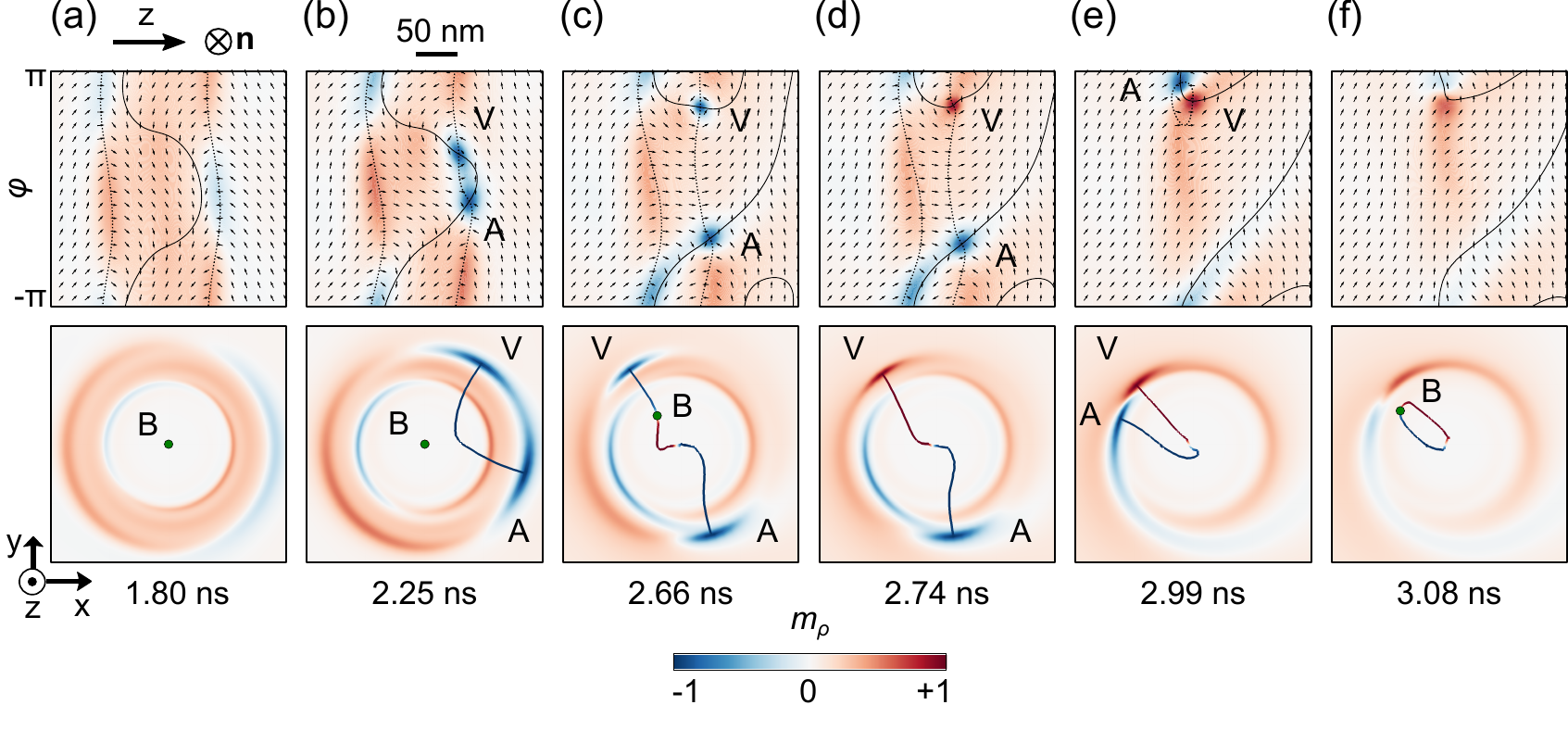}
\caption{\label{fig:sw_maps_mecanism} Snapshots illustrating the switching mechanism of BPW circulation in a permalloy wire with diameter~$\SI{90}{\nano\meter}$, for applied current amplitude $j=+\SI{1.2e12}{\ampere\per\meter\squared}$. (a)-(f) correspond to the time labels in Fig.\ref{fig:sw_thiele_energies_vs_time}(e). Top row: unrolled  maps of $m_\rho$ at the wire surface. \textbf{n} indicates the direction of the outer normal of the wire surface. Dashed lines correspond to $m_\varphi=0$ isovalues. Solid lines correspond to $m_z=0$ isovalues. Bottom row: same surface maps of $m_\rho$ seen from inside of the wire as a 3D view. Colored lines correspond to $m_\rho =1$ isovalues. Green dots show the position of the BP. A and V labels highlight surface vortex and antivortex, respectively.}
\end{figure*}

We describe here what becomes of a BPW when subject to an \OErsted field, depending on its initial circulation. In the present context we define the sign of circulation with respect to the $\unitvectvarphi$-axis, itself defined against the $\unitvect{z}$ direction [Fig.\ref{fig:schematics}]: for positive circulation $C+$ the azimuthal magnetization is parallel to the $\unitvectvarphi$-axis, and for negative circulation $C-$ it is antiparallel to $\unitvectvarphi$-axis. First we consider a head-to-head wall with no loss of generality, as a head-to-head domain wall and a tail-to-tail domain wall are equivalent through applying time-reversal and symmetry operation with respect to the $xy$ plane. In sections \ref{sec:switchingMechanism} and \ref{sec:topo} we nevertheless compare both. At this stage we disregard  spin-transfer effects, so that following inertia-related motion in the first stages of dynamics, the walls remain immobile after reaching their final configuration.

Unless otherwise stated, the simulations have been conducted using $\alpha=1$. This is an unphysically large damping, however suitable to describe quasistatic situations in a realistic sample, such as pulses of current with rise time of a few nanoseconds, relevant to our experimental situation\cite{bib-FRU2019b}. Therefore, we describe here a situation close to the minimum-energy path for magnetization processes. Considering a realistic damping value with sub-nanosecond rise time would induce complex ringing effects, like for precessional switching of macrospins\cite{bib-BAU2000c}. Finally, we performed simulations with both wires and thick-walled tubes, leading to negligible differences. Here, we illustrate the process with a permalloy wire with diameter $\SI{90}{\nano\meter}$.

Figure \ref{fig:sw_thiele_energies_vs_time} describes the behavior of the BPWs initially $C-$ or $C+$, while applying a positive current, thus favoring $C+$. Figure \ref{fig:sw_thiele_energies_vs_time}(a) qualitatively illustrates the rotation of magnetic moments in domains towards the \OErsted field at the wire surface and the evolution of the domain wall width up to the switching process. Figures \ref{fig:sw_thiele_energies_vs_time}(b)-(e) show the value over time of four quantities illustrating the process at play: (b)~the Thiele wall parameter, (c)~micromagnetic energies, the maxima and minima of the (d)~azimuthal and (e)~radial components of magnetization at the external surface. The BPW with initially positive circulation increases its width, reaching a plateau after about $\SI{0.5}{\nano\second}$~[Fig.\ref{fig:sw_thiele_energies_vs_time}(b)]. This is explained by the tilt of magnetization towards the azimuthal direction in the domains, thereby lowering the effective anisotropy inside the domain wall against the azimuthal direction, and thus increasing its width. The tilt reflects in the initial variation of $\max(m_\varphi)$ also evidenced for $C-$ [Fig.\ref{fig:sw_thiele_energies_vs_time}(d)], which is discussed in more detail in Appendix~\ref{sec-appendixCurrentModel}. The exchange energy increases as domains display a partial curling, the dipolar energy decreases as the head-to-head wall gets wider, and the Zeeman energy due to the \OErsted field decreases both in the domains and in the domain wall. The behavior of the BPW with negative circulation depends on the magnitude of the current. Below a critical current density $\jc$, the BPW contracts until it reaches a stable width~(not shown here). Above this threshold, the wall width decreases further until it reaches a minimum, before increasing rapidly towards the width of the BPW with positive circulation, all energies also coinciding~[Fig.\ref{fig:sw_thiele_energies_vs_time}(c)]. This suggests a reversal of circulation of the wall, confirmed by Fig.\ref{fig:sw_thiele_energies_vs_time}(d): in the initial state $\min(m_\varphi)=-1$ reflects the negative circulation, while $\max(m_\varphi)=0$ reflects magnetization in the domains. In the final state, \ie, after switching, $\max(m_\varphi)=+1$ now reflects the positive circulation, while $\min(m_\varphi)$ reflects the tilted magnetization in the domains. The variation of the radial component $m_\rho$ is by far more complex~[Fig.\ref{fig:sw_thiele_energies_vs_time}(e)], suggesting a non-trivial switching process, detailed in the next section.

\subsection{Switching of circulation of the BPW: microscopic mechanism}
\label{sec:switchingMechanism}

\begin{figure}[t]
\centering\includegraphics{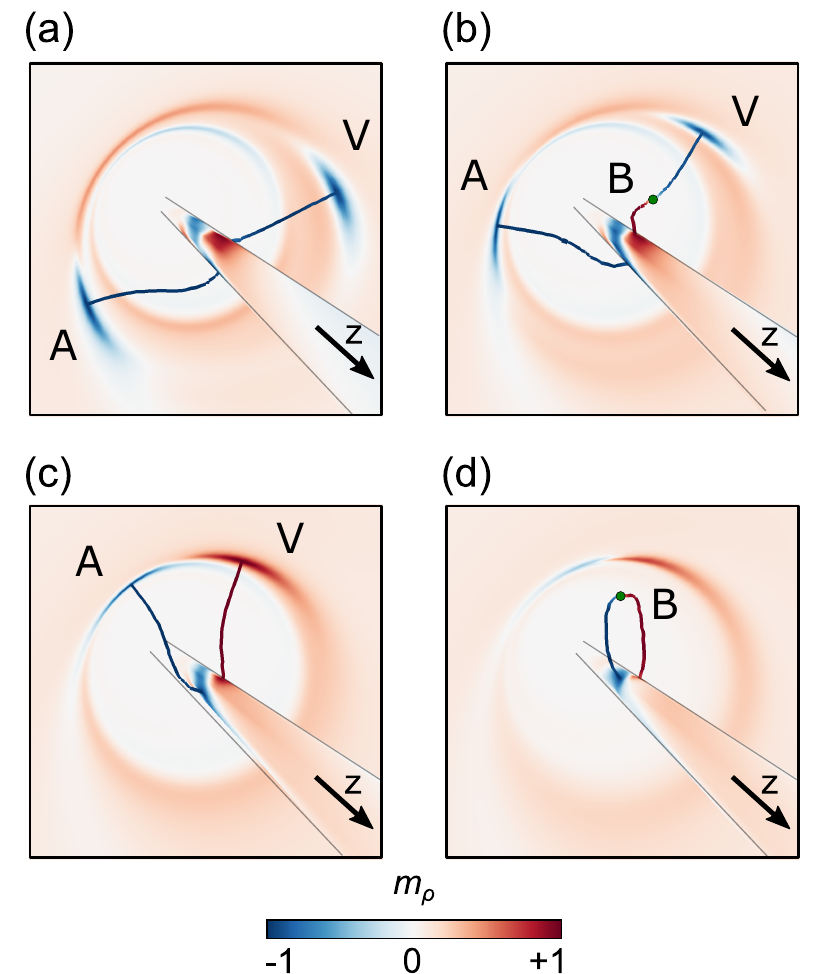}
\caption{\label{fig:sw_tube} Snapshots illustrating the switching mechanism in a permalloy tube with diameter~$\SI{90}{\nano\meter}$ for applied current amplitude $j=+ \SI{1.0e12}{\ampere\per\meter\squared}$. The $m_\rho$ colored maps seen from inside of the tube are the counterparts of Fig.\ref{fig:sw_maps_mecanism}. Colored lines correspond to $m_\rho =1$ isovalues.  Green dots show the position of the BP. A and V labels highlight surface vortex and antivortex, respectively.}
\end{figure}

The BPW texture displays the rotational symmetry at rest. This symmetry is not conserved through the switching process, which is far from a simple coherent rotation of the wall's inner degree of freedom. The nontrivial evolution in time of the out-of-plane magnetization component $m_\rho$ is illustrated  in Figure \ref{fig:sw_maps_mecanism}. To follow the magnetization transformation both at the wire surface as well as  in the volume we plotted the unrolled maps of $m_\rho$ on the external wire surface (top row) and the same $m_\rho$ surface maps seen from inside as a 3D view, completed with BP trajectories in the volume (bottom row).

Under the applied current, [Fig.\ref{fig:sw_maps_mecanism}(a)], magnetization in the domains rotates towards the azimuthal direction to follow the \OErsted field. Given the azimuthal rotation in the domains, the surface map has some similarity with a $\ang{180}$ domain wall in a thin film, made of a central micro-domain delimited by $m_\varphi=0$ isovalues (dashed back lines) and surrounded by two $\ang{90}$-like walls. The central micro-domain is characterized by an outward radial component ($m_\rho>0$), a well-documented fact\cite{bib-THI2006,bib-YAN2011b} for a wire at equilibrium and visible on Fig.\ref{fig:sw_thiele_energies_vs_time}(e) at $t=0$. Its sign results from positive magnetic charges of the head-to-head domain wall considered. On Fig.\ref{fig:sw_maps_mecanism}(a) an instability is developing, with locus of maximum or minimum of $m_\rho$ at $m_\varphi=0$. At these locations the torque due to \OErsted field is maximum as it is perpendicular to the local magnetization.The instability is accompanied with the deformation of the $m_z=0$ isoline (solid black line).  This behavior is consistent with the physics of walls in thin films, which tend to be of asymmetric Néel type or Bloch type to reduce the magnetostatic energy\cite{bib-HUB1998b}.

 When reaching locally $|m_\rho|=1$, the instability develops in a pair of vortex~(V) and antivortex~(AV)~ at the wire surface [Fig.\ref{fig:sw_maps_mecanism}(b)]. This corresponds to event b on Fig.\ref{fig:sw_thiele_energies_vs_time}(e), following a progressive decrease of $\min(m_\rho)$ and reflecting the rise of the instability. For topological reasons (discussed in the next section), these V and AV result from the continuous deformation of the ground state and thus share the same polarity (the same sign of $m_\rho$). The polarity happens to be negative, possibly because it allows to decrease the demagnetizing energy. Then, the V and AV move away one from another along the wire perimeter, which leaves in-between an area largely parallel to the magnetization direction in the domains~[Fig.\ref{fig:sw_maps_mecanism}(c)]. The phenomenon at play is clear: it is similar to a nucleation-propagation process, however not for an extended domain, but for the internal degree of freedom of a domain wall, such as the switching of the core of a vortex\cite{bib-THI2003}, or of the Néel cap in a Bloch wall\cite{bib-FRU2009b}.

The isolines $|m_\rho|=1$ in the 3D view allows to track the extent of the radial pocket inside the wire. From Fig.\ref{fig:sw_maps_mecanism}(b) to (c) (bottom row), it extends towards the axis, eventually reaching the existing Bloch point. After that, the Bloch point starts to move along this isoline towards the surface vortex, until it vanishes from the volume. The latter event [Fig.\ref{fig:sw_maps_mecanism}(d)] is accompanied with the change of the vortex polarity. In Fig. \ref{fig:sw_thiele_energies_vs_time}(e) this event correspond to the abrupt change in max$(m_\rho)$ due to the small size of the~BP, and the instantaneous character of a change of topology. At this stage the wall is of transverse-vortex type, for which the transformation back to the Bloch point under external stimulus is similar to the situation described in Ref.\cite{bib-FRU2015d}: the V and AV move further along the wire perimeter, until they merge~[Figs.\ref{fig:sw_maps_mecanism}(e),(f)]. This corresponds to event f on Fig.\ref{fig:sw_thiele_energies_vs_time}(e), which this time is associated with merge of V and AV of the same polarity and thus the creation of a BP. Finally, the new BP moves towards the center of the wire, ending in an immobile BPW with positive circulation.

We highlight below a few other features of the switching mechanism. First, it is similar in thick-walled tubes, except the lack of the BP in the volume originally ~[Fig.\ref{fig:sw_tube}(a)]. The latter implies the additional step of a BP nucleation by means of V/AV pair creation and transformation~[Fig.\ref{fig:sw_tube}(b)] at the inner surface. Then the BP travels towards the outer surface[Fig.\ref{fig:sw_tube}(c)].  Later a new one is created, travels towards the inner surface and annihilates[Fig.\ref{fig:sw_tube}(d)]. 

\begin{figure*}[t]
\centering\includegraphics[scale=1.0]{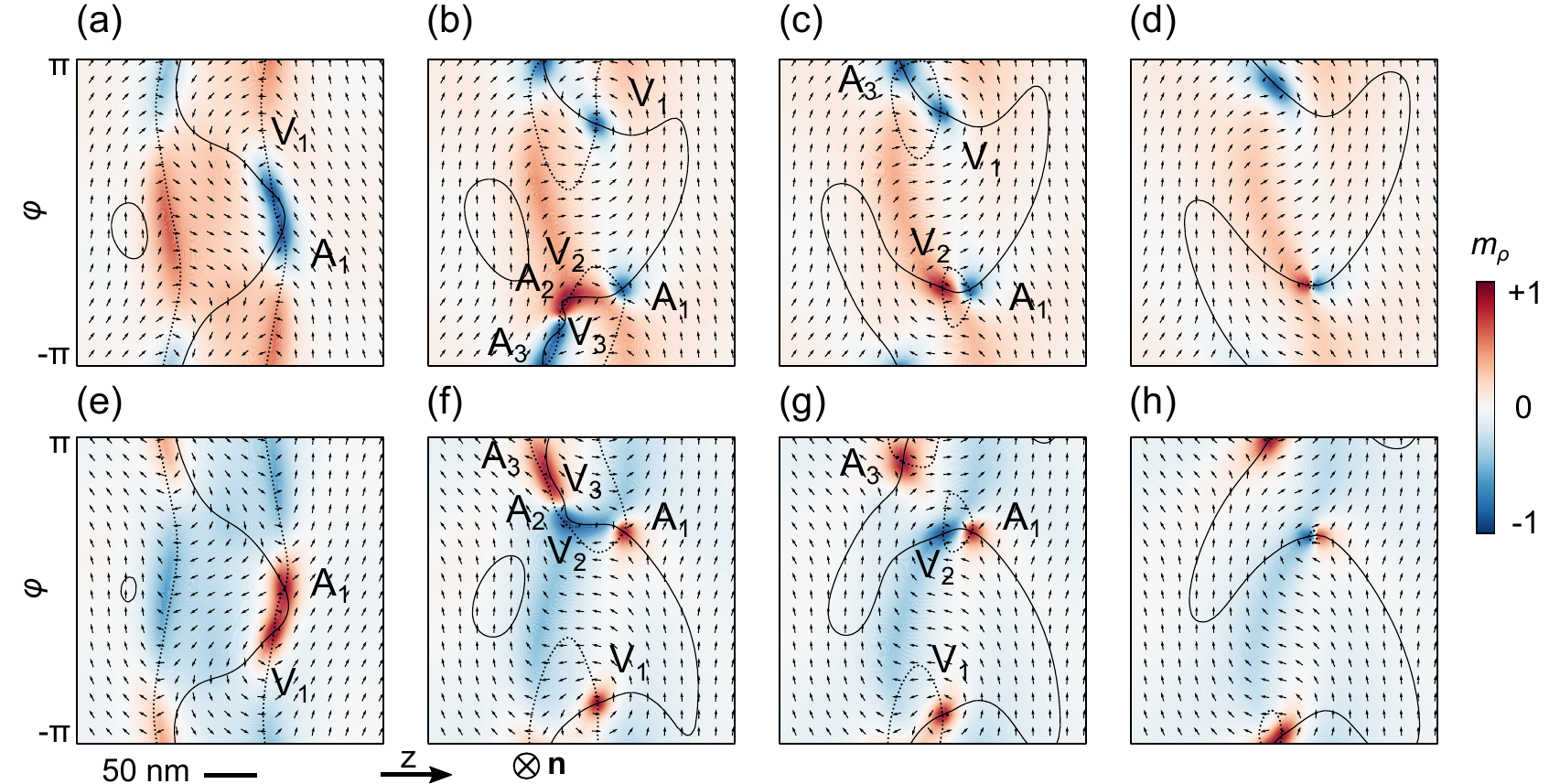}
\caption{\label{fig:sw_h2h_vs_t2t} Unrolled $m_\rho$ colored maps at the wire surface illustrating the switching mechanism in a permalloy wire with a head-to-head BPW (top row) and tail-to-tail BPW (bottom row). In both cases the circulation is initially negative $C-$, while the \OErsted field direction promotes positive circulation $C+$. Dashed lines correspond to $m_\varphi=0$ isovalues. Solid lines correspond to $m_z=0$ isovalues. A and V labels highlight surface vortex and antivortex, respectively. The normal \textbf{n} indicates the direction of the outer normal to the wire surface. All maps are plotted for the wire diameter~$\SI{90}{\nano\meter}$ and for applied current amplitude $j= +\SI{1.4e12}{\ampere\per\meter\squared}$, corresponding to an \OErsted field of \SI{39.5}{\milli\tesla} at the external wire surface.}
\end{figure*}

Second, the mechanism may become more complex for higher current densities, implying several pairs of V/AV. For example, Fig.\ref{fig:sw_h2h_vs_t2t}(a)-(d) show the case of a permalloy wire with diameter $\SI{90}{\nano\meter}$ and $j = \SI{1.4e12}{\ampere\per\meter\squared}$. The switching process now involves two pairs of V/AV. One pair interacts first with the BP on the axis, switching the polarity of the vortex. At the end of the process the V/AV pair with the same polarity does not nucleate a BP, while the V/AV with opposite polarity does, leading again to the same final state, a BPW with positive circulation.

Third, Figure \ref{fig:sw_h2h_vs_t2t}(e)-(h) also illustrates the equivalence of behavior of head-to-head and tail-to-tail domain walls, as expected. The situations displayed in the top and bottom rows of Fig.\ref{fig:sw_h2h_vs_t2t} are equivalent through applying two symmetry operations: time-reversal~(reversing both magnetization and applied current) and mirror symmetry around a plane containing the axis~(\eg, flipping top and bottom in the surface maps displayed). Accordingly, it can be checked that the top and bottom rows are equivalent under these two symmetries.

\subsection{Switching of circulation of the BPW: topological description}
\label{sec:topo}

In this section we analyze the switching process from a  topological point of view. To do so, we calculate the so-called \textit{winding numbers}, which measure the magnetization vector curling. This allows one to establish general features for the switching process.

 For isotropic spherical spins, parametrized as
$\textbf{m} = (m_1,m_2,m_3)$, the $S^2$-winding number reads \cite{Braun2012}, \cite{bib-BRA2018b}:
\begin{equation}
\label{eq:winding}
w  = \frac{1}{4\pi}\int_M \ve{m}\cdot \left(\partial_1 \ve{m}\times \partial_2 \ve{m}\right)dx_1dx_2,
\end{equation}
where $x_1$ and $x_2$ are arbitrary curvilinear
coordinates in real space, and $\partial_i=\partial/\partial x_i$. In Cartesian coordinates this expression
is often referred to as \textit{skyrmion number}. The manifold $M$ is usually understood
to be either the compactified plane $\mathbb{R}^2$ (V and AV in a thin film), or a 2-sphere (BP in the volume).

In practice, the application of Eq.\ref{eq:winding} to the surface texture with possibly a V or AV yields locally a half-integer number $w=q p/2$, where $q$ is usually referred to as the topological charge (or topological vorticity, or $S^1$-winding number), and $p$ the polarization \cite{bib-TRE2007}. V and AV are characterized by opposite topological charges: $q=+1$ and $q=-1$, correspondingly. The positive polarization $p=+1$ indicates the parallel alignment of the V/AV core with the outer normal and the negative $p=-1$ indicates the antiparallel alignement. The pair of V/AV with the same polarity has total $w=0$ and thus may be deformed continuously
into an uniform state. In the case of BP texture in volume, Eq.\ref{eq:winding} yields $w=\pm 1$, where positive sign indicates tail-to-tail BP type and negative sign head-to-head type.

In our study we consider topological objects in the volume (BP) and at the wire surface (V/AV). Strictly speaking, their winding numbers obtained with the same Eq.\ref{eq:winding} cannot be directly compared, and their sum should not obey any conservation law due to their contrasting geometrical nature. Conservation law may be established, for example, for purely flat nanomagnets which topological objects share the same manifold\cite{bib-TCH2005}. Nevertheless, the assessment of the total winding number change at the surface $w_{\textup{surf}}$ and, separately, in the volume $w_{\textup{vol}}$ reveals empirical rules for all switching processes described in this paper.

All snapshots in Figure \ref{fig:sw_maps_mecanism} may be classified into three topological situations: the BP is in the volume (a)-(c), the BP left the volume and caused a change of V polarity (d)-(e), the BP reenters the volume (f). Corresponding winding numbers are summarized in Table \ref{tab:case1}. 
\begin{table}[h]
\caption{\label{tab:case1} Winding numbers calculated for Figure \ref{fig:sw_maps_mecanism}.}
\begin{tabular}{|c||c|c|c|}
  \hline
  figure labels & (a), (b), (c)  & (d), (e)  & (f)\\
  \hline
  $\omega_{\textup{surf}}$ & $0$& $+1$ & $0$\\
  \hline
 $\omega_{\textup{vol}}$ & $-1$ & $0$ & $-1$\\
  \hline
\end{tabular}
\end{table}

The change in $w_{\textup{surf}}$ and $w_{\textup{vol}}$ between (c) and (d) events, as well as between (e) and (f) events obeys 
\begin{equation}
\label{eq:DeltaWinding}
\Delta \omega_{\textup{surf}}=\Delta \omega_{\textup{vol}}.
\end{equation}
This condition is also followed for thick-walled tubes [Fig.\ref{fig:sw_tube}] in the presence of inner and outer tube surfaces with normals pointing, correspondingly, towards negative and positive $\rho$-direction. Corresponding winding numbers are summarized in Table \ref{tab:case2}.  
\begin{table}[h]
\caption{\label{tab:case2} Winding numbers calculated for Figure \ref{fig:sw_tube}.}
\begin{tabular}{|c||c|c|c|c|}
  \hline
 figure labels & (a)& (b)  & (c)& (d)\\
   \hline
 $\omega_{\textup{surf}}^{\textup{in}}$+$\omega_{\textup{surf}}^{\textup{out}}$ & $0+0$& $-1+0$ & $-1+1$& $-1+0$\\
  \hline
 $\omega_{\textup{vol}}$ & $0$ & $-1$ & $0$& $-1$\\
  \hline
\end{tabular}
\end{table}

Moreover, the condition Eq.\ref{eq:DeltaWinding} is also satisfied in the case of multiple V/AV pairs formation, which happens with rising current amplitude. Figure \ref{fig:sw_h2h_vs_t2t} illustrates the situation for which the initial instability creates the first V/AV pair of V and AV of the same polarity similar to Fig.\ref{fig:sw_maps_mecanism} and further evolves towards a more complex texture with two extra V/AV pairs identified by $m_z=0$ and $m_\varphi=0$ isolines crossing. The shared polarity of each new V/AV pair is not necessarily the same as for the previously created pair. Moreover, in most cases we note that each additional pair has opposite polarization with respect to the previous pair, which is consistent with the hypothesis of the overall out-of-plane component minimization and the reduction of associated demagnetizing field penalty in the system. The corresponding winding numbers assessment is summarized in Table \ref{tab:case3}. 
\begin{table}[h]
\caption{\label{tab:case3} Winding numbers calculated for Figure \ref{fig:sw_h2h_vs_t2t}.}
\begin{tabular}{|c||c|c|c||c|c|c|c|}
  \hline
  figure labels & (a), (b)& (c) & (d)  & (e),(f) & (g) & (h) \\
   \hline
  $\omega_{\textup{surf}}$ & $0$& $+1$ & $0$ & $0$& $-1$& $0$\\
  \hline
  $\omega_{\textup{vol}}$ & $- 1$ & $0$ & $-1$ & $+1$& $0$& $+ 1$\\
  \hline
\end{tabular}
\end{table}

The situation looks equivalent for head-to-head and tail-to-tail BPWs, except that in the first case the BP interacts with a V by changing its polarity and in the second case with an AV in order to satisfy Eq.\ref{eq:DeltaWinding}. No matter how many intermediate V/AV pairs were created during the switching process, the starting V/AV creation and final V/AV annihilation events always follow the same pattern.     

The V/AV pair creation implies the additional exchange energy cost [Fig. \ref{fig:sw_thiele_energies_vs_time}(c)], and thus the threshold to be overcome for BP circulation switching. In the next section we calculate the corresponding critical current.

\subsection{Switching of circulation of the BPW: critical current density}%
\label{sec:Jcrit}

\begin{figure}[t]
\centering\includegraphics{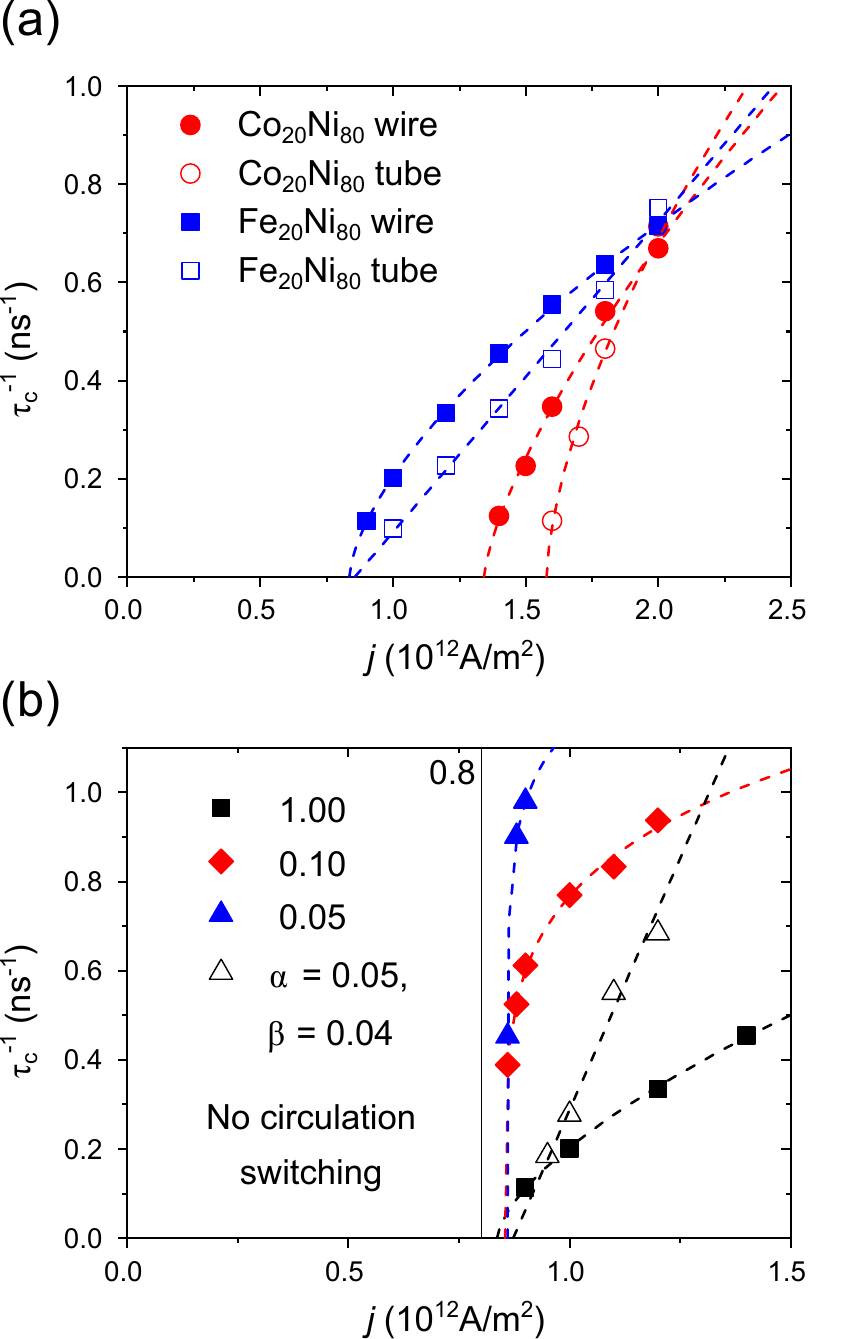}
\caption{\label{fig:tc_vs_J} Inverse switching time versus applied current density, for nanowires~(full symbols) or nanotubes~(open symbols) of external diameter $\SI{90}{\nano\meter}$ and composition either $\mathrm{Fe}_{20}\mathrm{Ni}_{80}$ or $\mathrm{Co}_{20}\mathrm{Ni}_{80}$. The dashed lines corresponds to the fit using the power law of Eq.\ref{eq:tcfit}. (a)~Comparison for wires and tubes, with $\alpha=1$. (b)~Comparison for several $\alpha$ values without and with spin-transfer torque effect for $\mathrm{Fe}_{20}\mathrm{Ni}_{80}$.}
\end{figure}

Here we are interested in describing the minimum current $\jc$ required to switch the circulation of a BPW, from initially antiparallel to parallel to the \OErsted field. The determination of a switching threshold against field or current is a delicate issue in numerical micromagnetism. Indeed, the computation time needs be finite in practice, so that a criterium is required to decide whether switching would not occur for a more extended time. A standard method to circumvent this difficulty is through performing a scaling of a parameter, for instance susceptibility below the threshold\cite{bib-RAV1998,bib-FRU2001}, or the switching time about the threshold\cite{bib-SUN2000b,bib-BED2010}. An interpolation through a few points and intercept with an axis then provides the threshold with high accuracy. In the present case, we consider the critical time $\tauc$ required for switching, above the threshold current.

The first step is to exhibit a criterium to define $\tauc$, as the complex and parameters-dependent dynamics revealed in section~\ref{sec:switchingMechanism} does not leave us with an ubiquitous one. After examination of various possibilities, the most robust choice turned out to be the time it takes for $\min(m_\varphi)$ at the external surface of the wire to change sign, directly highlighting the change of BPW circulation. In practice, the precise time for the change of sign is derived by fitting the curve in Fig.\ref{fig:sw_thiele_energies_vs_time}(d) using an $\mathrm{atanh}$ function.

The second step is to perform an interpolation, which requires a guess for the associated scaling law. A simple physical view is the following: the threshold current $\jc$ is associated with a critical slowing down of dynamics, and thus to the divergence of the characteristic time. A current density $j$ applied above the threshold suddenly brings the system out-of-equilibrium, giving rise to an effective field linear with $j-\jc$, to first order. The associated Kittel precessional frequency is expected to scale with this quantity, so that the switching time shall scale with $(j-\jc)^{-1}$. The inverse critical time indeed behaves fairly linearly versus the current density whatever the material or geometry, wire or tube~[Fig.\ref{fig:tc_vs_J}(a)]. The slight curvature arises probably because the Kittel's view for precession is too crude for the highly non-uniform process considered. To account for this curvature, in practice we used the phenomenological scaling law to fit these plots and extract precisely $\jc$:
\begin{equation}\label{eq:tcfit}
\tauc = \sigma_0(j-\jc)^{-p}
\end{equation}
In order to come closer to the experimental case, we considered more realistic damping parameters~$\alpha$, and also the effect of spin transfer, besides the \OErsted field~[Fig.\ref{fig:tc_vs_J}(b)]. The switching time is largely affected by these parameters, however the threshold current $\jc$ is not. This shows that the present results remain valid even for current pulses with a sub-ns rise time, and that the \OErsted field is indeed the crucial and largely dominating reason for the switching of circulation.

\begin{figure}[]
\centering\includegraphics[width=87mm]{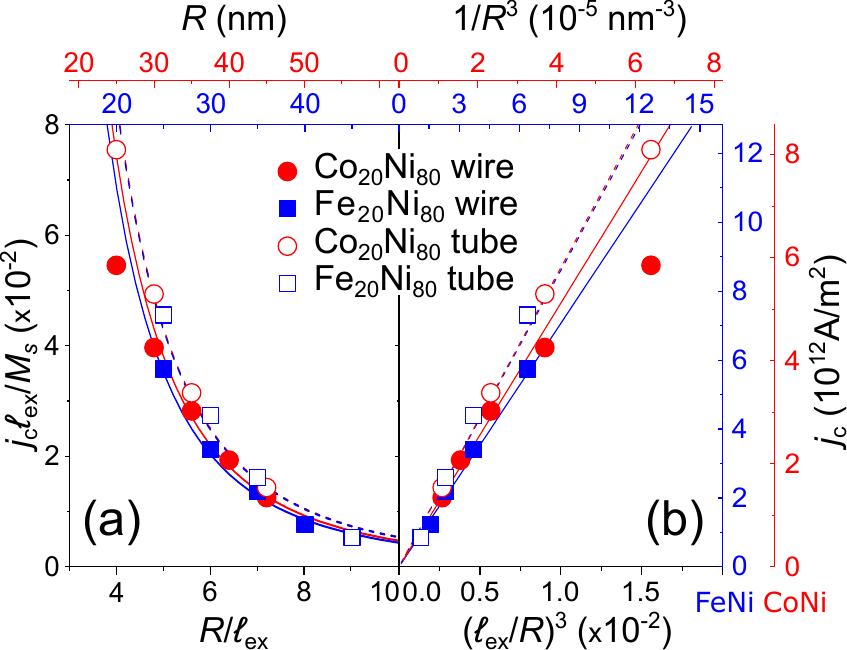}
\caption{\label{fig:tc_vs_R}Critical current density $\jc$ in nanowires versus (a)~their radius~$R$ and (b)~$1/R^3$. The dashed lines (tubes) and solid lines (wires) correspond to  Eq.\ref{eq:scalingVersusRadius}. For tubes, $C$ = 10.73 and 10.82 for  $\mathrm{Fe}_{20}\mathrm{Ni}_{80}$ and $\mathrm{Co}_{20}\mathrm{Ni}_{80}$ respectively. For wires, $C$ = 8.81 and 9.54 for  $\mathrm{Fe}_{20}\mathrm{Ni}_{80}$ and $\mathrm{Co}_{20}\mathrm{Ni}_{80}$ respectively. For Bottom and left axis display dimensionless quantities, while top and right axis provide real quantities for the two material considered, $\mathrm{Fe}_{20}\mathrm{Ni}_{80}$ and $\mathrm{Co}_{20}\mathrm{Ni}_{80}$.}
\end{figure}

Finally, to provide a complete view of the switching process, we evaluated the threshold current $\jc$ against the radius~$R$, illustrated on Fig.\ref{fig:tc_vs_R}(a) based on tubes. Plotting $j_c$ against $1/R^3$ reveals a close-to-linear law~[Fig.\ref{fig:tc_vs_R}(b)]:
\begin{equation}\label{eq:scalingVersusRadius}
\jc \approx C \frac{A}{\muZero\Ms R^3},
\end{equation}
with $C$ a dimensionless coefficient. This scaling law is supported by an analytical model balancing exchange with \OErsted Zeeman energy in the wire geometry~(Appendix~\ref{sec-appendixCurrentModel}), and partly by dimensional analysis~(Appendix~\ref{sec-appendixCurrentDimensionalAnalysis}). Let us comment the impact of this result. First, Eq.\ref{eq:scalingVersusRadius} is valid for any magnetically soft material, upon the proper scaling of length and current density~(Appendix~\ref{sec-appendixDimensionless}):
\begin{equation}%
\label{eq:scalingVersusRadiusDimensionless}
\tilde{j}_\mathrm{c}= (C/2)r^{-3},
\end{equation}
with $r=R/\lex$ the dimensionless radius. This law predicts a switching current only $\SI{20}{\%}$ larger than the experimental one. We consider that this is a fair agreement, considering error bars on exchange stiffness, material composition, and possible sample defects or the role of temperature during the current pulse in the experiments. Second, the switching current reaches experimentally unpractical values below $R\approx\SI{30}{\nano\meter}$~(circa $6\lex$). This means that the investigations published previously and neglecting the \OErsted field remain valid in the low-radius regime, the circulation of BPWs tending to switch positive with the direction of wall motion\cite{bib-THI2006,bib-YAN2011b}, thus with the flow of electrons and hence negative with the direction of current. Also, there must exist a threshold regime where the effect of the \OErsted field and the chirality of the LLG equation compete, leading to an unpredictable circulation and wall speed. Conversely, for large radius note that the azimuthal tilt of magnetization in the domains scales with the same $1/R^3$ law~(see Appendix~\ref{sec-appendixCurrentModel}). This means that circulation switching should have no limit for large radius, occurring always for the same wall angle, around $\ang{270}$ following Fig.\ref{fig:sw_thiele_energies_vs_time}.

\begin{figure}[]
\centering\includegraphics[width=87mm]{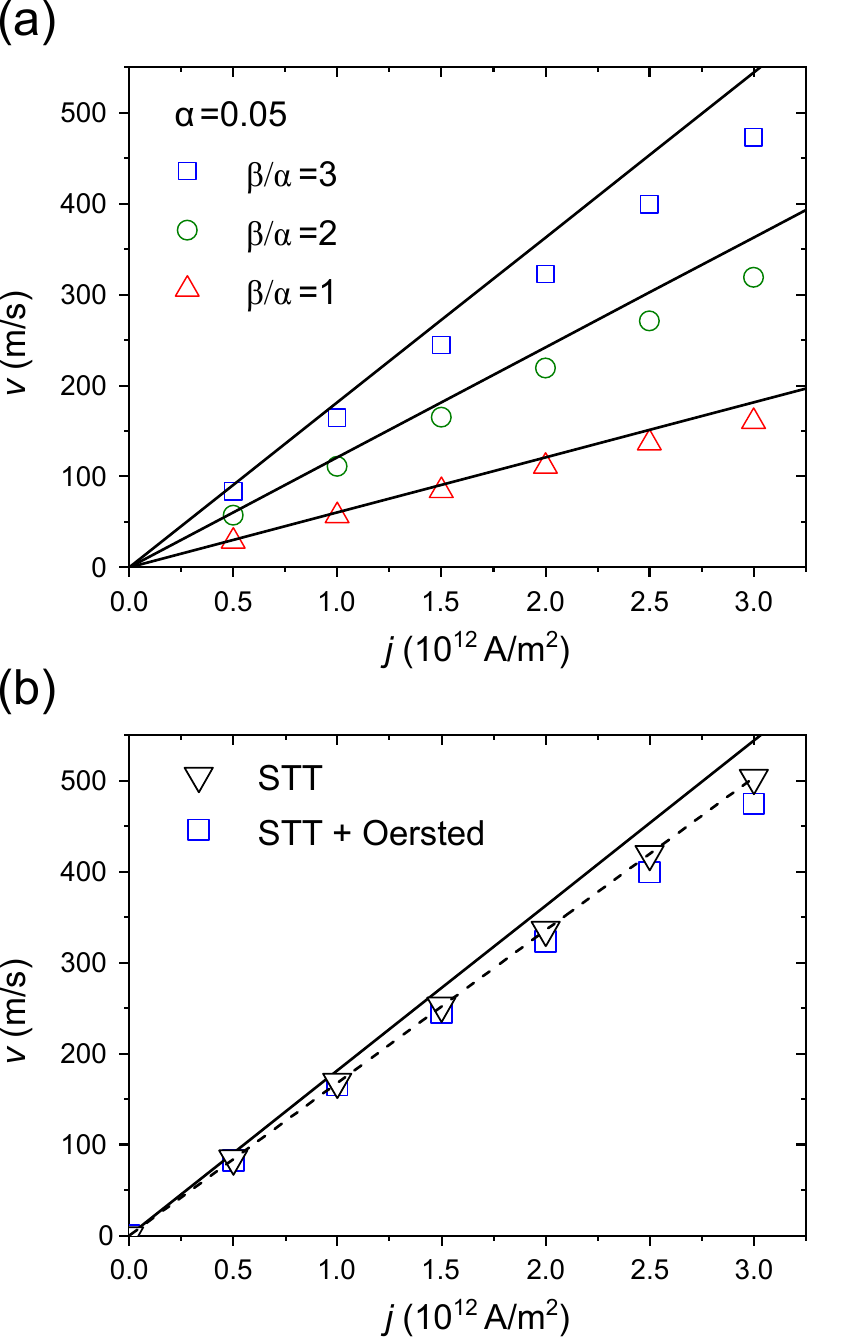}
\caption{\label{fig:speed} BPW speed vs applied current density for $\mathrm{Co}_{20}\mathrm{Ni}_{80}$ thick-walled tube with outer diameter $\SI{90}{\nano\meter}$ and inner diameter $\SI{10}{\nano\meter}$. Solid black lines correspond to Eq.\ref{eq:velocity}. (a) Both spin-transfer and \OErsted effects are considered for different values of $\beta/\alpha$ and $\alpha = 0.05$. (b) Comparison between purely spin-transfer case and spin-transfer together with \OErsted field case for $\alpha = 0.05$ and $\beta = 0.15$. The dashed line is a linear fit of spin-transfer only case.}
\end{figure}

\section{Spin-transfer-driven motion of the BPW under \OErsted field}
\label{sec-speed}

The \OErsted field proves crucial in experiments to stabilize BPWs, and it selects a sign of circulation of the BPW opposite to the one expected from the chirality of the LLG equation during motion, if the current is large enough\cite{bib-WIE2010}. Yet, the \OErsted field does not break the symmetry between the two domains, and thus cannot drive a steady-state motion, for which spin-transfer torque remains crucial. Here we bring together the two effects, to elucidate to which extent the speed predicted so far based on spin-transfer alone, remains relevant. As discussed in \ref{sec:thick-walled}, to avoid the numerical pinning of the BPW subjected to electrical current we substitute the wire by a thick-walled tube of the same diameter with a PBPW. Figure \ref{fig:speed}(a) plots the PBPW speed $v$ as a function of applied current, including spin-transfer and \OErsted field effects. For simplicity we start directly with the PBPW circulation favorable to the \OErsted field to avoid the circulation switching. For current amplitudes relevant experimentally, the PBPW dynamics obeys the steady regime equation 
\begin{equation}%
\label{eq:velocity}
\ve{v}=\frac{\beta}{\alpha}\ve{u}.
\end{equation}
We attribute the discrepancies between Eq.\ref{eq:velocity} and micromagnetics  at high speeds to usual numerical artifacts related to the energy over-dissipation\cite{bib-ALO2014}. As expected, we did not observe any signature of Walker breakdown and did not find any significant change in velocity related to the presence of the \OErsted field, as illustrated in Fig.\ref{fig:speed}(b).

\section{Conclusion}

We reported recently experimentally the key role played by the \OErsted field in magnetically soft cylindrical nanowires, to stabilize Bloch-point walls~(BPWs) and reach speed $>\SI{600}{\meter\per\second}$ through spin-transfer torque\cite{bib-FRU2019b}. Here we used micromagnetic simulations, analytical modeling and topological arguments to understand in detail and quantitatively the underlying phenomena, in particular the switching from negative to positive circulation of the BPW, with respect to the applied current. The key result is the $1/R^3$ dependence of the switching threshold, with $R$ the wire radius, with the effect of the \OErsted field becoming predominant for wire radius above typically $\SI{30}{\nano\meter}$. To the contrary, the speed of walls remains largely determined by spin-transfer alone, in a below-Walker regime. Thanks to a generalized micromagnetic scaling of lengths and densities of current, the present result can be applied to wires made of any magnetically soft material.

\section{Acknowledgements}


The project received financial support from the French National Research Agency (Grant No. JCJC MATEMAC3D). M. S. acknowledges a grant from the Laboratoire d’excellence LANEF in Grenoble (ANR-10-LABX-51-01).

\appendix

\section{Dimensionless micromagnetics with \OErsted field}\label{critical current model}
\label{sec-appendixDimensionless}

It is a standard procedure to scale lengths to the dipolar exchange length in the micromagnetics of soft magnetic materials, so that results are material independent. Here, we extend this scaling to include the current density $\vect j$, source of the \OErsted field. The volume density of micromagnetic energy reads:
\begin{equation}
E = A\left({\vectNabla\cdot\vect m}\right)^2-\frac12\muZero\vectM\cdot\vectHd -\muZero\vectM\cdot\vectHOE
\label{eq:micromagneticEnergy}
\end{equation}
In Eq.\ref{eq:micromagneticEnergy}, $\vectHOE=(j \rho/2)\unitvecttheta$, with $\rho$ the distance to the wire axis. To switch to dimensionless variables, we normalize Eq.\ref{eq:micromagneticEnergy} with the dipolar constant $\Kd=\muZero\Ms^2/2$, turning to dimensionless energy~$e$. Simultaneously, we normalize lengths with the dipolar exchange length $\lex$, the magnetization vector with spontaneous magnetization~$\Ms$, turning into unity~$\vect m$, magnetic fields with spontaneous magnetization~$\Ms$, written~$\vect h$. These normalizations are the usual ones for soft magnetic materials. In the present case, we also normalize the volume density of charge current~$\vect j$ with $\Ms/\lex$, written $\tilde{\vect j}$. Eq.\ref{eq:micromagneticEnergy} becomes:
\begin{equation}
e = \left({\vectNabla_u\cdot\vect m}\right)^2-\vectm\cdot\vecthd -\tilde{\vect j}\tilde{\rho}\,\vectm\cdot\unitvecttheta
\label{eq:micromagneticEnergyNorm}
\end{equation}
$\vectNabla_u$ stands for the gradient operator against the dimensionless coordinates~$u$, and $\tilde{\rho}$ is the dimensionless distance to the axis. Thus, the results of our manuscript are valid for any soft magnetic material, provided that the above normalization is used. We drew a number of figures in the manuscript based on these dimensionless variables [Figs.\ref{fig:static_slice_thielevsR}, \ref{fig:tc_vs_R}].

\section{Critical current: dimensional analysis}
\label{sec-appendixCurrentDimensionalAnalysis}

Numerical simulations reported in Sec.\ref{sec:Jcrit} have shown that the threshold current $\jc$ required to switch the circulation of a BPW scales with $R^{-3}$, $R$ being the radius of the nanowire. Here, we discuss the physical meaning of this scaling law, based on dimensional analysis.

The switching depends on the balance between different energies, related to the exchange interaction, the demagnetizing field and the \OErsted field. These involve the following physical quantities: exchange stiffness $A$ in \si{\joule\per\meter}, the dipolar constant $\Kd=(1/2)\mu_0\Ms^2$ in \si{\joule\per\meter\cubed}, and the Zeeman energy involving $\mu_0\Ms$.
Thus, the relevant independent physical quantities that may be involved in determining $\jc$ are: the exchange stiffness $A$, the spontaneous magnetization $\Ms$, the vacuum permeability $\mu_0$ and the nanowire radius $R$. Therefore, an expansion of the law determining $\jc$ must necessarily be:
\begin{equation}
\jc \sim A^\alpha \; R^\beta \; \mu_0^\gamma \; \Ms^\delta\;,
\label{eq:jcDimensional}
\end{equation}
with $\alpha$, $\beta$, $\gamma$ and $\delta$ dimensionless coefficients to be determined. This equation can be translated into its SI units:
\begin{multline}
\si[per-mode=reciprocal]{\ampere\per\square\meter}=\left({\si[per-mode=reciprocal]{\meter\kilo\gram\per\second\squared}}\right)^\alpha
  \cdot \si{\meter\tothe{\beta}}
  \cdot \\ \left({\si[per-mode=reciprocal]{\meter\kilo\gram\per\second\squared\per\ampere\squared}}\right)^\gamma
  \cdot \left({\si[per-mode=reciprocal]{\ampere\per\meter}}\right)^\delta\;,
\label{eq:jcUnits}
\end{multline}
which leads to the following set of equations, related to the powers of meter, kilogram, second and Ampere:
\begin{eqnarray}
  \alpha + \beta + \gamma - \delta &=& -2 \label{eq:jcUnitPowers1} \\
  \alpha + \gamma &=& 0 \label{eq:jcUnitPowers2} \\
  -2\alpha -2\gamma &=& 0 \label{eq:jcUnitPowers3} \\
  -2\gamma +\delta &=& 1 \label{eq:jcUnitPowers4}
\end{eqnarray}
Eq.\ref{eq:jcUnitPowers2} and Eq.\ref{eq:jcUnitPowers3} are equivalent, so that this set becomes:
\begin{eqnarray}
  \beta &=& 2\alpha-4 \label{eq:jcUnitPowers7} \\
  \gamma  &=& -\alpha \label{eq:jcUnitPowers8} \\
  \delta &=& 1-2\alpha \label{eq:jcUnitPowers9}
\end{eqnarray}
This set of equations is under-determined once, with $\alpha$ taking any possible value. Writing $\alpha=1+n$, we end up in:
\begin{equation}
\jc\sim\sum_n C_n \frac{A}{\mu_0\Ms R^3}\;\left({\frac{R}{\lex}}\right)^{2n}
\end{equation}
with coefficients $C_n$. So, dimensional analysis alone does not allow to explain that $\jc\sim1/R^3$, which corresponds simply to a predominant $C_0$. This suggests that the dipolar exchange length is largely irrelevant. Said differently, the remaining term $A/(\mu_0\Ms R^3)$ can be decomposed as the ratio of $A/R^2$ with $\mu_0\Ms R$, suggesting a competition of exchange energy and \OErsted Zeeman energy alone, to determine the switching of circulation. A model based on this competition is detailed below.

\section{Critical current: analytical model for the scaling law}
\label{sec-appendixCurrentModel}

\begin{figure}[h]
\centering
\includegraphics{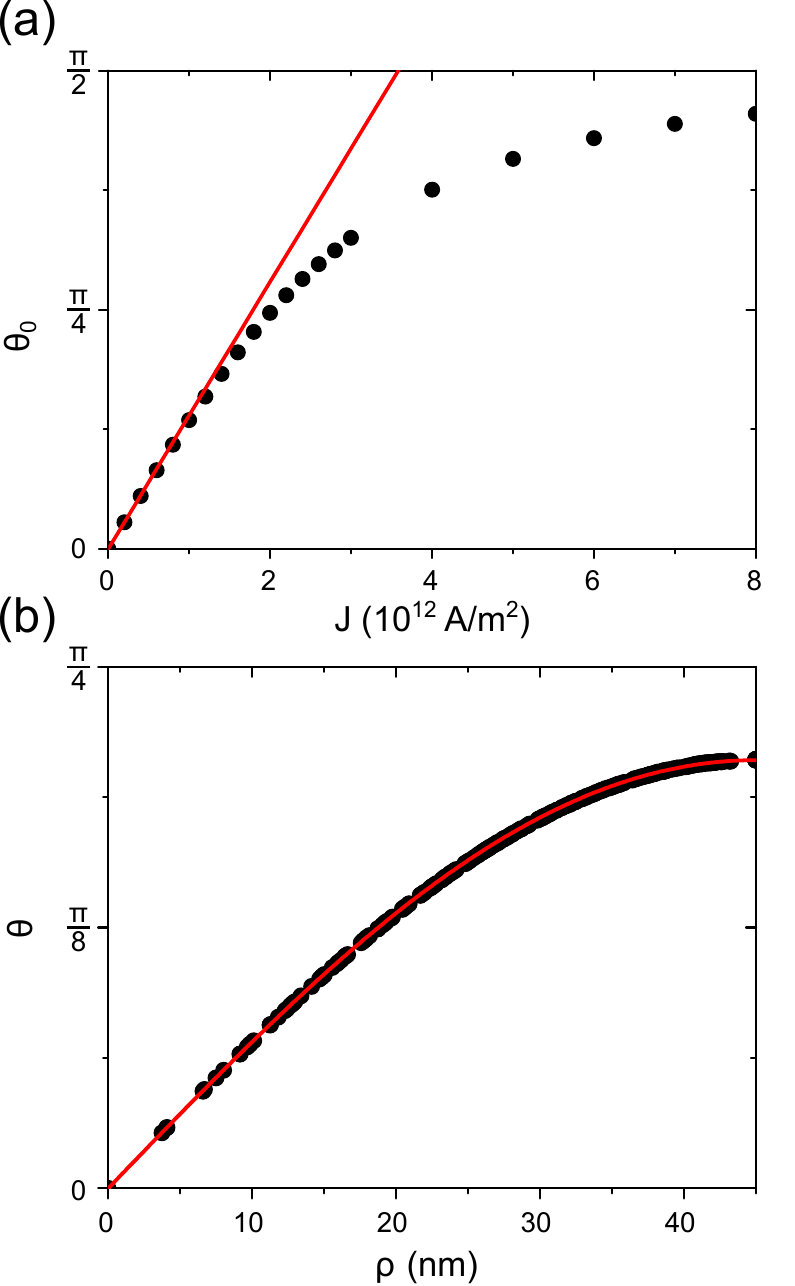}
\caption{Comparison of the analytical model versus micromagnetic simulations, for the effect of the \OErsted field on magnetization in an extended domain. We used following parameters: wire radius $R=\SI{45}{\nano\meter}$, $A=\SI{1.1E-11}{\joule\per\meter}$ and $\Ms=\SI{6.7E5}{\ampere\per\meter}$. The line stand for the (linear) analytical solution~[Eq.\ref{eq:jversusthetaDomainModel}], while symbols stand for micromagnetic simulations. (a)~Current density $j$ required to reach a given azimuthal tilt of magnetization $\theta_0$ on the nanowire surface, in an initially uniformly-magnetized domain. (b)~Radial variation of the tilt of magnetization~$\theta$, for $j=\SI{1.6E12}{\ampere\per\squared\meter}$.}
\label{fig:thetavsr}
\end{figure}

Here we propose a simple argument to explain the $1/R^3$ dependence of the threshold current~$\jc$ for circulation switching~[Fig.\ref{fig:tc_vs_R}]. The model does not intend to be a rigorous one, however to put forward the physical ground responsible for this scaling law.

The previous section suggested that $\jc$ is predominantly determined by the competition of exchange and Zeeman \OErsted energies. We consider this competition in the curling effect in the domains, for which the absence of dipolar fields, and the translational symmetry, allow a straightforward modeling. The \OErsted field forces magnetization at radius $\rho$ to acquire an azimuthal component, tilting from $\unitvect{z}$ towards $\unitvectvarphi$. We propose to describe this tilt with a test function:
\begin{equation}
\theta(\rho) = \theta_0 \sin{\left(\frac{\pi}{2}\frac{\rho}{R}\right)}.
\end{equation}
$R$ is the external radius and $\theta_0 = \theta(\rho = R)$. The volume density of exchange energy $E_\mathrm{ex}$ and of the Zeeman \OErsted energies, read:
\begin{eqnarray}
  E_{\mathrm{ex}} &=& A \left[\left(\frac{\partial{}\theta{}}{\partial{}\rho}\right)^2 +  \frac{\sin^2{\left(\theta{}\right)}}{\rho^2} \right], \\
  E_{\mathrm{Z}} &=& -\mu{}_0M_{\mathrm{s}}\frac{j\rho}{2}\sin{\left(\theta\right)}.
\end{eqnarray}

The total energy for a wire length $L$ is:
\begin{equation}
\label{eq:totalEnergyDomainModel}
\mathcal{E}_\mathrm{T} = \int_{0}^{R} 2 \pi{} \rho L  \left(   E_{\mathrm{ex}}    +   E_{\mathrm{Z}}    \right)  \mathrm{d}\rho.
\end{equation}
This integral may be evaluated by making use of a Taylor series expansion of $\sin(x)$ around $x=0$, and consideration of the test function for $\theta(\rho)$. Expanding to second order for $\theta_0$, Eq.\ref{eq:totalEnergyDomainModel} becomes:
\begin{multline}
\xi_{\mathrm{T}} = \frac{\pi{}^3 LA \theta_0^2}{4} \left( 2-  \frac{\pi^2}{6} + \frac{10\pi^4}{1728} \right)\\-\frac{ \pi^2 \mu_0 M_{\mathrm{S}} L j R^3 \theta{}_0  }{8}\left(   1-\frac{\pi^2}{36}  \right).
\end{multline}
By minimizing the total energy with respect to $\theta_0$, we find a simple relation between $j$, $\theta_0$ and $R$:
\begin{equation}
\label{eq:jversusthetaDomainModel}
j = \frac{4\pi A\theta_0}{\mu_0 \Ms R^3} \frac{\left( 1-\frac{\pi^2}{36}  \right) }{\left( 2-  \frac{\pi^2}{6} + \frac{10\pi^4}{1728}   \right)}.
\end{equation}

Fig.\ref{fig:thetavsr}a compares the relationship between $\theta_0$ and $j$, the linear law in the present analytical model~[Eq.\ref{eq:jversusthetaDomainModel}], and micromagnetic simulations. The match is excellent at low angle, until $\theta_0\approx\pi/4$, a range consistent with the expansion of the $\sin$ function. The deviation beyond this point is not troublesome, as the associated density of current is too high to be experimentally relevant. The perfect match between the simulation and the analytical model is equally observed in the radial dependence of the tilting of magnetization~[Fig.\ref{fig:thetavsr}(b)] for low applied current densities. In conclusion, Eq.\ref{eq:jversusthetaDomainModel} is perfectly valid, suggesting the origin of the law $\jc\sim1/R^3$ as resulting from the dominant competition between exchange and Zeeman energies.


\bibliographystyle{apsrev4-2}
\bibliography{BPW_biblio}
\end{document}